\documentclass[conference]{IEEEtran}

\IEEEoverridecommandlockouts

\usepackage{amsmath}
\usepackage[final]{graphicx}

\renewcommand{\theenumi}{\arabic{enumi}}

\renewcommand{\theenumii}{\alph{enumii}}

\renewcommand{\theenumiii}{\arabic{enumiii}}

\makeatother

\begin{document}

\title{GossiCrypt: Wireless Sensor Network Data Confidentiality
Against Parasitic Adversaries
\thanks{$^{\star}$ Jun Luo and Panagiotis Papadimitratos are
equally contributing authors.}
\author{
  \IEEEauthorblockN{Jun Luo$^{\star}$}
  \IEEEauthorblockA{Department of Electrical and Computer Engineering\\
    University of Waterloo\\
    Waterloo, Ontario, Canada N2L 3G1\\
    Email: \texttt{j7luo@engmail.uwaterloo.ca}}
\and
  \IEEEauthorblockN{Panagiotis Papadimitratos$^{\star}$~~~~~~~~~Jean-Pierre Hubaux}
  \IEEEauthorblockA{School of Computer and Communication Sciences\\
    EPFL (Swiss Federal Institute of Technology in Lausanne)\\
    CH-1015, Lausanne, Switzerland\\
    Emails: \texttt{\{panos.papadimitratos~~~~~~~}\\
    \texttt{~~~~~~~jean-pierre.hubaux\}@epfl.ch}}
}}
\maketitle

\begin{abstract}
Resource and cost constraints remain a challenge for wireless sensor
network security. In this paper, we propose a new approach to
protect confidentiality against a \textit{parasitic adversary},
which seeks to exploit sensor networks by obtaining measurements in
an unauthorized way. Our low-complexity solution,
\textbf{GossiCrypt}, leverages on the large scale of sensor networks
to protect confidentiality efficiently and effectively. GossiCrypt
protects data by symmetric key encryption at their source nodes and
re-encryption at a randomly chosen subset of nodes \emph{en route}
to the sink. Furthermore, it employs key refreshing to mitigate the
physical compromise of cryptographic keys. We validate GossiCrypt
analytically and with simulations, showing it protects data
confidentiality with probability almost one. Moreover, compared with
a system that uses public-key data encryption, the energy
consumption of GossiCrypt is one to three orders of magnitude lower.
\end{abstract}

\section{Introduction} \label{sec:in}

Wireless sensor networks (WSNs) have been an active field of
research over the last few years, with a number of technical issues
largely resolved. Onwards wider adoption, security becomes
increasingly important and, eventually, security mechanisms a
prerequisite~\cite{Perrig04}. Numerous significant efforts have been
made along this line, including public-key cryptography (e.g.,
\cite{Tinypk,Sizzle}) as the means to digitally sign messages and
establish symmetric keys, as well as symmetric-key based encryption
and authentication for improved efficiency (e.g.,
\cite{Perrig02,Karlof04}). However, \emph{sensor data
confidentiality} has been largely overlooked to this date. Ensuring
that sensor-collected data are accessed only by authorized entities
has been viewed mostly as a secondary concern.

Encrypting data at their source sensor node, with a symmetric key
shared with the sink, is a straightforward confidentiality mechanism.
However, it does not fully address the problem at hand. An adversary
can actively exploit the poor physical protection of nodes, as it
would be too costly and thus unrealistic to make them
tamper-resistant. It is relatively easy for an adversary to
physically access the node memory contents~\cite{Hartung05}, and
extract the symmetric key used for data encryption. Such an attack is
vastly simpler than a cryptanalytic one against the key. In fact, the
adversary could progressively compromise keys of numerous nodes, and
eventually be able to decrypt a significant fraction of, if not all,
data produced by the WSN.

We are concerned with sensor data confidentiality in such a setting,
where cryptographic keys can be physically compromised. We focus on a
novel type of adversary we term \emph{parasitic}: it seeks to
\textbf{exploit} a WSN, e.g., deployed for scientific measurements,
industrial (mining, oil) field data, or even patients' health data
collection, rather than disrupt, degrade, or prevent the WSN
operation. A parasitic adversary, defined in detail in
Sec.~\ref{sec:adv}, aims at obtaining measurements with the least
expenditure of own resources, and the least disruption of the WSN it
``attaches'' itself to. Essentially, the longer the symbiotic
relation of the adversary with a fully functioning WSN remains
unnoticed, the more successful the parasitic adversary will be.

One naive solution against (symmetric) key compromise is to let
sensors encrypt each outgoing measurement with the public key of the
sink. As long as the sink is not compromised, it is the only one
able to decrypt those message and the parasitic adversary is
thwarted. However, software implementations of public-key
operations, albeit computationally feasible, consume energy
approximately \textbf{three orders of magnitude} higher than
symmetric key encryption~\cite{PKLife}. Hardware implementations of
public key encryption (PKE) can significantly reduce energy
consumption, but they remain accordingly costlier than symmetric key
encryption (SKE) hardware implementations (Sec.~\ref{sec:ee}).

Therefore, we are facing the challenge of protecting data
confidentiality against parasitic adversaries in an energy efficient
manner. To this end, we propose here \textbf{GossiCrypt}, whose
mechanisms are tailored to and leverage on the salient features of
WSNs. GossiCrypt comprises two building blocks: (i) a probabilistic
\emph{en route re-encryption} scheme, with the source node always
encrypting the data and with relaying nodes \emph{en route} to the
sink flipping a coin to ``decide'' whether to perform re-encryption,
and (ii) a \emph{key refreshing} mechanism that installs new
sensor-sink shared symmetric keys to selected nodes.

Key refreshing is the immediate response to the compromise of a
cryptographic key, but it can mitigate such an attack only to a
certain extent: it is hard for the WSN operator to infer which keys
were compromised. Also, running a network-wide key distribution
protocol frequently can be very costly in an energy-constrained
environment. More important, within two refreshing events, the
adversary would still be fully capable to decrypt data from nodes
whose keys were compromised. This is where the \emph{en route}
re-encryption complements our (infrequent) key refreshing: data (or
keys) can be decrypted by the adversary only if all the keys used for
source and en-route encryption are compromised.

GossiCrypt has extremely simple key management requirements and very
low complexity operation. Each sensor shares one data encryption
symmetric key with the network sink. In addition, a single parameter
drives probabilistically the participation of each node in en-route
encryptions. This simplicity is inherent in gossiping protocols,
with nodes flipping a coin to determine, e.g., if they should
synchronize their databases or relay a message~\cite{Demers87,
Haas02}. This inspires the name of our scheme, as the decision is on
(re-)encrypting rather than on relaying a packet. Key refreshing is
also simple, as it is performed with randomly chosen nodes. Overall,
simplicity renders GossiCrypt broadly applicable.

Our main contribution is an efficient and highly effective, as our
evaluation shows, scheme to ensure sensor data confidentiality. The
objectives of GossiCrypt are specified in Sec.~\ref{sec:qp}. We
validate the effectiveness of our scheme analytically and
experimentally. Attacked by a parasitic adversary that continuously
compromises new nodes to obtain their encryption keys, GossiCrypt
protects the confidentiality of data with probability almost one. At
the same time, the comparison with PKE shows that the GossiCrypt
energy expenditure is significantly lower. Another contribution is
the introduction of the parasitic adversary, a realistic type of
attacker for a wide range of commodity and tactical WSNs. To the best
of our knowledge, this is a novel yet realistic and highly effective,
unless thwarted, type of adversary.

In the rest of the paper, we first provide the system and adversary
models. Then, we present an overview of our scheme and present in
detail its constituent protocols. In Sec.~\ref{sec:pe}
and~\ref{sec:sim}, we analyze our scheme and provide an experimental
validation. Due to space limitation, literature survey is omitted; a
detailed discussion of related work can be found in
\cite{tech-report}. We discuss a number of issues related to our
scheme in Sec.~\ref{sec:disc}, and conclude in Sec.~\ref{sec:con}.

\section{System Model}\label{sec:sys_model}

The WSN comprises $N$ \emph{sensor nodes}, each with a unique
identity $S_i$, and a \emph{network sink} $\Theta$ performing data
collection and key refreshing. It is straightforward to consider
multiple sinks, even with distinct roles, yet we omit this for
simplicity in presentation. Each node $S_i$ shares a symmetric key,
$K_{i,\Theta}$, with the sink, and knows the public key,
$PuK_\Theta$, of the sink. The sink is equipped with all
$K_{i,\Theta}$.

Beyond these \emph{end-to-end}, sensor-to-sink, associations, nodes
may share symmetric keys with their neighbors, to enable link-layer
security primitives (e.g., TinySec~\cite{Karlof04}). However, such
security mechanisms are beyond the scope of this work and they can
clearly coexist with our scheme.

We describe the data of interest with the help of two parameters,
$T$ and $\delta$; the user seeks to collect data:
\begin{itemize}
\item From a fraction $0< \delta \le 1$ of the WSN nodes,
\item Over a period of $T$ seconds, for each node $S_j$, for $j=1,\ldots,\lceil\delta N\rceil$.
\end{itemize}

The actual values of $T$ and $\delta$ can vary. $T$ can range from a
short period, $t_0$, for a single sensor measurement, to a
sufficiently long period for a comprehensive measurement collection.
In general, $T=kt_0$, with $k>0$ an integer. Similarly, $\delta=1/N$,
i.e., targeting at a certain node, may be meaningful, but in practice
$\delta$ will be a significant fraction of $N$.\footnote{WSNs
deployed for (often one-time) event detection (e.g., forest fire or
bridge structural faults) would correspond to $\delta=1$, and $T$
equal to the period from the WSN deployment to the event/alarm
occurrence.} We do not dwell on the exact measurement extraction
method, which can be performed in many ways orthogonal to our scheme.

We assume that $N$ ranges from hundreds to thousands, as, for
example, in WSNs for commercial inventory, habitat monitoring,
industrial and mining field data, and geological measurements.
Experience from prior deployments, with node placement sparser than
the monitored physical system and relatively long history of
measurements necessary to capture the studied phenomena, teaches that
data sensed by each and every node is significant. This implies that
in-network data aggregation is not an option in such deployments; we
assume this is the case in this work. We also assume WSNs enabling
applications that do not undergo development. Thus, the entire
operating system (apart from certain tunable parameters) is stored in
\textit{read-only memory} (ROM). Finally, WSN nodes are not
tamper-resistant or store cryptographic keys in tamper-resistant
components, due to cost considerations.

\section{Adversary Model}\label{sec:adv}

We identify a new type of adversary we term \textit{parasitic}. Its
objective is to exploit deployed wireless sensor networks, by
accessing in an unauthorized manner data collected by those WSNs.
More specifically, a parasitic adversary:

\begin{enumerate}
\item Seeks to obtain the WSN data collected according to the parameters $\delta$ and
$T$.

\item Can be physically present, at each point in time, \textbf{only} at a much smaller fraction of the
area covered by $\lceil\delta N\rceil$ sensor nodes.

\item Can physically access data stored at sensor nodes and retrieve their cryptographic keys.

\item Can be mobile~\cite{Ostrovsky91}, i.e., compromise different sets of nodes over different
time intervals. ``Mobile" traditionally refers to virtual moves (in
terms of compromising system entities); here, it also represents
physical moves of the adversary.

\item Can compromise in the above-described manner at most one sensor per $\tau$ seconds.
We assume $\tau \ll T$.
\end{enumerate}

\begin{figure}[t]
   \begin{center}
        \includegraphics[width=0.7\columnwidth]{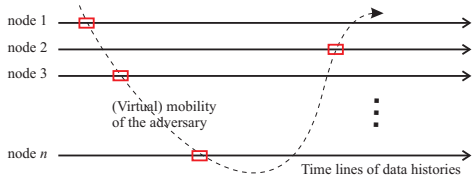}
    \caption{Mobility of the parasitic adversary.}\label{fig:MobileAdv}
   \end{center} %\vspace{-3ex}
\end{figure}

The characteristics of the parasitic adversary reflect its realism.
Constrained presence (assumption~2) is meaningful, because,
otherwise, the adversary could deploy its own WSN and trivially
obtain the data the WSN user collects (assumption~1). It exploits
obvious weaknesses of WSNs (assumption~3): poor physical protection
makes it relatively easy to obtain data encryption
keys~\cite{Hartung05}. The parasitic adversary is
\textit{unobtrusive}, that is, cannot modify the implemented
protocols stored in ROM (Sec.~\ref{sec:sys_model}). Furthermore, it
can utilize its resources intelligently. Mobility (assumption~4),
illustrated in Fig.~\ref{fig:MobileAdv}, shows that the adversary
can be in the proximity of different nodes for periods of time
during which it either compromises the node, or obtains snapshots of
their measurement histories, or intercepts messages sent from nodes
within its receiving range.
%% check the last phrase of this paragraph

The strength of the adversary is evident from assumption~5: the time
needed to physically compromise a single node, albeit significant if
nodes are carefully designed, is much shorter than $T$, the period
over which data are to be collected. In other words, the benefit of
the adversary from compromising sensor nodes is far reaching. %Note
%however that compromising a key does not imply the adversary
%automatically obtains the node's measurements.
The adversary could remain within range of the compromised node and
trivially intercept all its transmissions. But such an attack would
be self-defeating: from assumption~2, the adversary would certainly
capture much less than $\lceil\delta N\rceil$ measurements. From a
different point of view, assumption~2 captures the difficulty to
deploy a network of eavesdroppers within one hop of all previously
compromised nodes. The eavesdroppers' transceivers would need to be
highly sensitive (and thus more expensive than that of a sensor
node) to cover a meaningful fraction of the targeted WSN. Overall,
leaving ``sentry'' nodes behind would be comparable to deploying a
WSN by the adversary.
%% Check what I commented out - to improve the flow, and also because the statement sounded gossicrypt specific.

We assume that the protocol design and implementation are such that
remote node compromise is prevented. For example, the adversary
cannot exploit arbitrary software weaknesses and make a sensor node
disclose its cryptographic keys. Such robustness should be possible
given the relatively simple functionality of WSN node software,
compared to that of more complex systems (e.g., desktop or portable
computers). We also assume that the sink \emph{cannot} be
compromised by the adversary. Readers are referred to \cite{Zhang05}
for the investigations on compromise of low-end mobile sinks.
Moreover, denial-of-service (DoS) attacks, including jamming in
various protocol layers~\cite{Wood02}, Sybil/Node replication
attacks~\cite{Parno05}, or ``wormhole'' formation~\cite{Papad08} are
beyond the scope of this work: countermeasures to those attacks can
coexist with our protocols. Neither do we consider physical
destruction of WSN nodes, which would not benefit the adversary.

\section{GossiCrypt} \label{sec:qp}

GossiCrypt aims at ensuring confidentiality, that is, preventing any
unauthorized access to data collected by a WSN. It does not seek to
protect data coming from every single sensor, but rather intends to
fulfill the following property, for some protocol-specific constant
$0<\Delta<1$:

\vspace{1mm} \noindent \textbf{$\Delta_T-$Confidentiality:}
\emph{Data collected from a WSN comprising $N$ nodes are
$\Delta_T-$confidential if the adversary cannot obtain all
measurements performed by more than $\lceil N\Delta\rceil$ sensor
nodes over a given time interval $T$.} \vspace{1mm}

This is a \emph{safety} property, i.e., a property related to a
system-specific unwanted situation: obtaining measurements from a
given fraction of sensor nodes over a period of time, meaningful
with respect to the system and application, is prevented. In
Sec.~\ref{sec:sa} we will show that GossiCrypt satisfies this
property against parasitic adversaries with probability almost one.

We emphasize that GossiCrypt does not seek to provide sensor data
authenticity and integrity. The reason is that if a key is
compromised, an adversary (not necessarily a parasitic one) can
impersonate the corresponding sensor and inject fabricated messages.
Nonetheless, data that originate from non-compromised nodes have
their authenticity and integrity protected. We also clarify that
GossiCrypt does not seek to hide the identities of sensor nodes,
achieve data source untraceability, or satisfy any notion of
anonymity, unlinkability, or privacy. Clearly, confidentiality
relates to privacy, but, again, all GossiCrypt seeks to provide is
the \emph{confidentiality} of the data provided by sensor nodes.

\subsection{Data Encryption}\label{sec:pl}

\begin{figure}[t]
   \begin{center}
        \includegraphics[width=0.8\columnwidth]{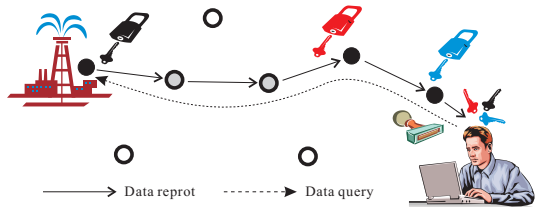}
    \caption{Securing data collection with GossiCrypt and
    query authentication ($\mu$TESLA [32] for example).}\label{fig:GossiCrypt}
   \end{center} %\vspace{-3ex}
\end{figure}

We distinguish sensor nodes into two types, \emph{data sources} and
\emph{relaying nodes}, with each node assuming either role at
different points in time. We denote by $\mathit{GossiCrypt}_E$ the
data encryption operation of GossiCrypt. As illustrated in
Fig.~\ref{fig:GossiCrypt}, it is executed by nodes on the path from
a data source to a sink (inclusive), with the outcome (i.e.,
re-encrypting or not) at each relaying node being random (with
probability {\boldmath$q$}).

The path may be one hop, if the sink is within the transmission range
of the sensor node, but this is not cost-effective; in general, the
sink is at a distance of multiple hops from data source(s). The path
discovery is orthogonal to $\mathit{GossiCrypt}_E$. It can be
determined by a (secure) routing protocol, for example, forming an
authenticated tree rooted at the sink~\cite{Perrig02}, possibly
on-the-fly, as a result of the query sent out from a sink.
$\mathit{GossiCrypt}_E$ can be employed on top of any path discovery
protocol and does \emph{not} impose extra requirements. For the rest
of the discussion, we assume that, minimally, each $S_i$ knows the
next node towards $\Theta$ on a $path_{S_i,\Theta}$ without the
transmitted packet carrying the routing information.

For a sensor measurement $m$, a symmetric key $K_{i,\Theta}$ shared
by $\Theta$ and $S_i$, a message authentication code
$MAC(K_{i,\Theta},\ldots)$, and $q \in (0,1)$ the protocol-specific
parameter governing the \emph{en route} re-encryption,
$\mathit{GossiCrypt}_E(K_{i,\Theta},$ $path_{S_i,\Theta}, q, m)$ is
invoked by $S_i$ acting as a source:

\begin{enumerate}
\item \textbf{Source node, $S_i$:}
\begin{enumerate}
 \item Generate a nonce $n$ for the communication with sink $\Theta$.
 \item Calculate $H = MAC(K_{i,\Theta}, m, n, S_i)$.
 \item Encrypt $m, n, H$ with $K_{i,\Theta}$ to obtain ciphertext
       $\sigma_i=\{m, n, H\}_{K_{i,\Theta}}$.
 \item Transmit packet $p_i=\sigma_i, S_i$ to the first relaying node $S_j$ on $path_{S_i,\Theta}$.
\end{enumerate}

\item \textbf{Relaying node, $S_j$:}
\begin{enumerate}
\item Upon receipt of a packet $p_i$, generate a random number $x\in[0,1]$.
      If $x>q$, relay $p_i$ to the next relaying node $S_k$ on
      $path_{S_i,\Theta}$, or to $\Theta$. Otherwise,
\item Generate ciphertext $\sigma_j = \{p_i\}_{K_{j,\Theta}}$.
\item Append own identity $S_j$ to $\sigma_j$.
\item Relay packet $p_j=\sigma_j, S_j$ to the next relaying node $S_k$
      along $path_{S_j, \Theta}$, or to $\Theta$.
\end{enumerate}

\item \textbf{Sink $\Theta$:}
\begin{enumerate}
\item Upon receipt of a packet $p_k$, retrieve $K_{k,\Theta}$, the key shared
      with $S_k$, and decrypt $\sigma_k$. If the source, $S_i$, cleartext $m, n, H$, is obtained, go
      to (c). Otherwise,
\item Obtain ciphertext $\sigma_l$ and $S_l$. Decrypt $\sigma_l$ with $K_{l,\Theta}$. Repeat successively
      for all $S_l$ that re-encrypted the packet, till obtaining the source clear-text $m, n, H$.
\item \label{itm:vn} Determine if $n$ was previously seen. If so, discard the packet. Otherwise,
\item Compute $H'=\mathit{MAC}(K_{i,\Theta}, m, n, S_i)$. Discard the packet if $H'\ne H$.
      Otherwise, deliver $m$ to the WSN user.
\end{enumerate}
\end{enumerate}

\subsection{Key Refreshing}\label{sec:kr}

To defend against the progressive compromise of an increasing number
of nodes, $K_{i,\Theta}$ keys should be refreshed, i.e., replaced
with new $K'_{i,\Theta}$ keys. The sink is typically unaware of
which nodes are already compromised. Thus, it selects randomly an
$S_i$ node to refresh, among a set of $N' \leq N$ nodes. This
selection is, in general, made among the data source nodes of
interest (the $\delta$ fraction of $N$ as defined in
Sec.~\ref{sec:sys_model}), and all the intermediate nodes that
connect those sources to the sink. In other words, the refreshing
effort focuses on the same part of the network that is meaningful
for the adversary to target.

Given a particular system design for the nodes, it is not very
difficult to have an arguably pessimistic estimation of the rate of
physical node compromise, as per Sec.~\ref{sec:adv}. Then, based on
this estimate of $\tau^{-1}$, the \emph{key refreshing rate}
$\lambda_r$ can be selected accordingly by the sink, and conveyed to
all nodes via an authenticated control message. Confidentiality of
$\lambda_r$ is not needed, as the adversary would, at best,
compromise nodes at its maximum possible rate $\tau^{-1}$.
Authenticity, however, is clearly required, to ensure that an active
adversary does not ``slow down'' the key refreshing.

Symmetric-key based key transport techniques, similar to those
in~\cite{ISO}, are effective only if the adversary, having
previously compromised $K_{i,\Theta}$, cannot intercept the key
refreshing protocol messages. Moreover, an interactive key
establishment protocol, for example, initiated by the sink, would
reveal the identity of the node whose key is being refreshed. The
adversary could eavesdrop all messages sent and received from the
sink, and hence gain a significant advantage: that is, know which
nodes were refreshed and then re-compromise them.

To thwart these two vulnerabilities, we propose a key refreshing
protocol with two variants. This is essentially a key transport
protocol; but it leverages on (i) the $\mathit{GossiCrypt}_E$
operation, with optional public key encryption at the source sensor
node, and (ii) the integration of the key refreshing with the data
collection. As a result, the key refreshing protocol is similar to
the data encryption protocol, presented in Sec.~\ref{sec:pl}. There
are two main differences: a random point process generator
\cite{Bremaud99}, $\mathit{RGen}(\lambda_r)$, used to generate (key
refreshing) events with intensity $\lambda_r$, and a $\mathit{flag}$
set to indicate to the sink that a new key $K'_{i,\Theta}$ is
included in the message (which, otherwise, externally appears
identical to any measurement/data reporting message). The protocol
operates as follows:

\begin{enumerate}
\item \textbf{Source node, $S_i$:}
\begin{enumerate}
 \item Upon an event of $\mathit{RGen}(\lambda_r)$, generate a new key
 $K'_{i,\Theta}$; wait for the time till the next data report.
 \item Upon a data report to be returned, delay the report to be combined with the next one, and generate a nonce $n$ for the communication with sink $\Theta$.
 \item Calculate $H = MAC(K_{i,\Theta}, \mathit{flag}, K'_{i,\Theta}, n, S_i)$.
 \item Encrypt $\mathit{flag}, K'_{i,\Theta}, n, H$ with $K_{i,\Theta}$, to obtain ciphertext
       $\sigma_i=\{\mathit{flag}, K'_{i,\Theta}, n, H\}_{K_{i,\Theta}}$.
 \item Transmit packet $p_i=\sigma_i, S_i$ to the first relaying node $S_j$ on $path_{S_i,\Theta}$.
\end{enumerate}

\item \textbf{Relaying node, $S_j$:} \\ Identical to the operation
for $\mathit{GossiCrypt}_E$ (Sec.~\ref{sec:pl}).

\item \textbf{Sink $\Theta$:}
\begin{enumerate}
    \item Perform the steps (3).(a)-(b) as specified in Sec.~\ref{sec:pl}, to obtain
     the source, $S_i$, cleartext $\mathit{flag}, K'_{i,\Theta}, n, H$.
    \item Determine if $n$ was previously seen. If so, discard the packet. Otherwise,
    \item Calculate $H' = MAC(K_{i,\Theta}, \mathit{flag}, K'_{i,\Theta}, n, S_i)$. If $H' \ne H$, discard the packet.
          Otherwise, replace $K_{i,\Theta}$ with $K'_{i,\Theta}$.
    \end{enumerate}
\end{enumerate}

The protocol installs a new key even if the adversary intercepts the
message \emph{en route} to the sink, unless the adversary is
physically within one hop from the previously compromised and now
to-be-refreshed $S_i$. In the later case (which is rare due to the
constrained physical presence of an adversary), the adversary can
decrypt the message and obtain the key. To prevent this, we propose
the following variant of the above key refreshing protocol:

\begin{enumerate}
\item \textbf{Source node, $S_i$:} Identical to the above key refreshing operation, with
the additional step between (b) and (c), and replacing
$K'_{i,\Theta}$ with $\sigma\kappa_i$ afterwards:
\begin{enumerate}
 \renewcommand{\theenumii}{\alph{enumii}$^+$}
 \setcounter{enumii}{1}
 \item Encrypt $K'_{i,\Theta}$ with $PuK_{\Theta}$, the public key of the
 \renewcommand{\theenumii}{\alph{enumii}}
       sink, and obtain the ciphertext $\sigma\kappa_i=\{S_i,K'_{i,\Theta}\}_{PuK_{\Theta}}$.
\end{enumerate}

\item \textbf{Relaying node, $S_j$:} \\ Identical to the operation
for $\mathit{GossiCrypt}_E$ (Sec.~\ref{sec:pl}).

\item \textbf{Sink $\Theta$:}
Identical to the above key refreshing operation, with the additional
step:
\begin{enumerate}
    \setcounter{enumii}{3}
    \item Decrypt $\sigma\kappa_i$ with $PrK_{\Theta}$, the private key of the sink, and check if the
    obtained node identity is $S_i$. If so, replace $K_{i,\Theta}$ with $K'_{i,\Theta}$.
    \end{enumerate}
\end{enumerate}

This second variant's use of PKE resembles mechanism 1 of the ISS/IEC
11770-3 standard~\cite{ISO-2}. It ensures that even in the unlikely
event the adversary is within one hop of the refreshed node, still,
it cannot obtain the new $K'_{i,\Theta}$. The only option for the
adversary would be to re-compromise $S_i$.

\section{Protocol Analysis} \label{sec:pe}

We analyze the security level of GossiCrypt and also compare its
energy expenditure with a possible alternative in this section. Our
security analysis focuses only on the parasitic adversary; further
discussion on other adversaries is given in Sec.~\ref{sec:disc} and
\cite{tech-report}. The security analysis applies to both data
encryption and key refreshing (with or without PKE) protocols, as
they follow the same principle.

\subsection{Security Analysis} \label{sec:sa}

In this section, we describe a model of GossiCrypt and evaluate it
against the $\Delta_T-$Confidentiality property (Sec.~\ref{sec:qp})
and the parasitic adversary (Sec.~\ref{sec:adv}). Our analysis,
accompanied by simulation results in Sec.~\ref{sec:sim}, shows that
even with a significant fraction of sensor nodes compromised,
GossiCrypt safeguards confidentiality \textbf{with probability
almost one}.

Fundamental for the analysis is the fraction of \emph{correct},
i.e., not compromised, nodes; this is determined by the behaviors of
the sink refreshing and the adversary compromising keys. Therefore,
we model the \emph{state} of the system, the number of correct
nodes, as a stochastic process. Our security analysis on GossiCrypt
is based on the stationary regime of this process.

Since the sink cannot in general know which keys are already
compromised, a randomized strategy on selecting which node to refresh
is a reasonable choice. We assume that the sink does so with an
effective\footnote{The model covers the two options (with or without
PKE) of the key refreshing protocol described Sec.~\ref{sec:kr}.
Although the key refreshing without PKE might allow the adversary to
obtain the new key, it is still highly possible that new keys are not
exposed to the adversary, as the adversary cannot be ubiquitously
present (also pointed out in~\cite{Anderson04}). Thus, the model
still applies but with the refreshing rate $\lambda_r$ discounted by
a factor.} refresh rate $\lambda$. Recall that the sink governs the
selection procedure through setting the parameter $\lambda_r$. The
adversary, compromising nodes at rate $\tau^{-1}$, is also modeled as
selecting the next node to compromise (or to test if the key was
refreshed)\footnote{A model that assumes the rate of testing
differing from that of compromising does not fundamentally change the
stationary distribution.} arbitrarily. This is so, because the
adversary is also in general unaware of which keys were refreshed by
the sink.\footnote{In a static sink network, the adversary might
gradually, over a long period of eavesdropping, infer (part of) the
communication paths connecting the sensor nodes to the sink. This
could allow the adversary to launch a deterministic attack (e.g.,
starting from the sink's neighbors and then moving outwards,
compromising their upstream nodes). This might allow the adversary to
fight back against symmetric-key based refreshing if and only if it
has compromised the entire path connecting the refreshed node to the
sink. However, this attack would be completely ineffective against a
public key based refreshing (as described in Sec.~\ref{sec:kr}). The
only approach that could allow the adversary to detect if some node
$S_k$ re-encrypted a message with a new key (that does not allow the
adversary to decrypt the message and then can guide its
re-compromise), would be to intercept the message before it is
received and after it is relayed by $S_k$. But this would imply
physical presence of the adversary along the entire path and
eventually the source node(s). This would contradict assumption~2.
Therefore, the deterministic, targeted compromise pattern would be
essentially impossible and thus pointless, and thus no more effective
than a random one. We note that it is also possible that the sink
counters deterministic attack patterns with similarly structured
refresh patterns. However, investigation of those albeit interesting
is not provided here due to space limitations. For example, the
efficiency of the scheme could greatly enhanced if the public key
refreshing protocol is run with nodes near the sink, to ``break''
chains of fully compromised paths and make symmetric-key refreshing
effective even against this deterministic attack.} Although an
adversary physically close to a source node $S_i$, may detect a
key-refreshing, its physical presence is limited to a negligible
fraction of the network. Note that re-encryption deprives the
adversary from this ability elsewhere. The aforementioned assumptions
suggest that both the sink and the adversary follow Markov chains
\cite{Bremaud99} in choosing the next target. In particular, the
adversary may follow a deterministic trajectory, which is a special
Markov chain with deterministic transitions.

The system size depends on the behavior of the sink. If the sink is
static and the data collection paths change slowly, if at all, over
time, both the sink and the adversary could have a clear view on
which nodes they need to target: the source sensor nodes of interest
and the relaying nodes en-route to the sink. Or better even, from
the adversary's point of view, the slightly smaller subset of
sources and relaying nodes en-route to the point it intercepts the
measurement packets. As a result, the system is this known subset of
nodes with size $N'<N$. On the other hand, if a mobile sink is
used~\cite{Kansal04,Titra04,Luo06}, the adversary cannot predict the
data collection paths. This results in a larger system size, which
essentially can be all nodes, offering higher robustness against the
adversary at the expense of complexity in operating the mobile sink.
We emphasize however that our analysis is applicable to both cases.
All one needs to do is to view $N$ below as the effective system
size.

We assume that the times of performing refreshing and compromising
can be modeled as two independent Poisson processes with intensities
$\lambda$ and $\tau^{-1}$ respectively. We also assume that, at each
time point in the processes, either the sink approaches a node and
refreshes it or the adversary captures a node and compromises it, no
matter whether the node has been compromised or not. The Poissonian
and independence assumptions are not essential. The easily drawn
analogies between our model and the teletraffic models
\cite{Bonald06} imply that the stationary distribution is insensitive
to all other characteristics beyond the intensities.
%% check the ".... are not essential." seems ok, but just to be sure.

Based on these assumptions, we describe the system states as a
continuous Markov chain $\{X(t)\}_{t\geq0}$ driven by the Poisson
processes. Since such a chain is characterized by its subordinated
chain $\{\hat{X}_{n}\}_{n\geq0}$ \cite{Bremaud99}, we focus on this
discrete Markov chain. A direct observation on the system is that the
more numerous the compromised nodes, the less the efficiency of the
adversary (thus the higher the efficiency of the sink) is and vice
versa. The reason is clear: when many nodes are compromised, the
probability of fruitlessly re-compromising becomes high. This reminds
us of the celebrated model described by Paul and Tatiana Ehrenfest
(sometimes referred to as The Urn of Ehrenfest)~\cite{Ehrenfest90}
for understanding the diffusion through a porous
membrane.\footnote{The model can be briefly described as follows
\cite{Bremaud99}: there are $N$ particles that can be either in
compartment $A$ and $B$. Suppose at time $t$, there are $i$ particles
in $A$. The diffusion process behaves as if someone chooses a
particle at random and moves it to another compartment at time $t+1$.
Therefore, the transition probability is
$p_{ij}=\frac{i}{N}~(j=i-1)\mathrm{,
or}~\frac{N-i}{N}~(j=i+1)\mathrm{, or}~0~\mathrm{(otherwise)}$.} The
system we consider differs from the Urn of Ehrenfest in that the
``self'' transition probability is non-zero (i.e., $p_{ii}>0$) and
also that the transition probability depends on the rates $\lambda$
and $\tau^{-1}$.

\begin{figure}[htb]
   \begin{center}
        \includegraphics[width=0.85\columnwidth]{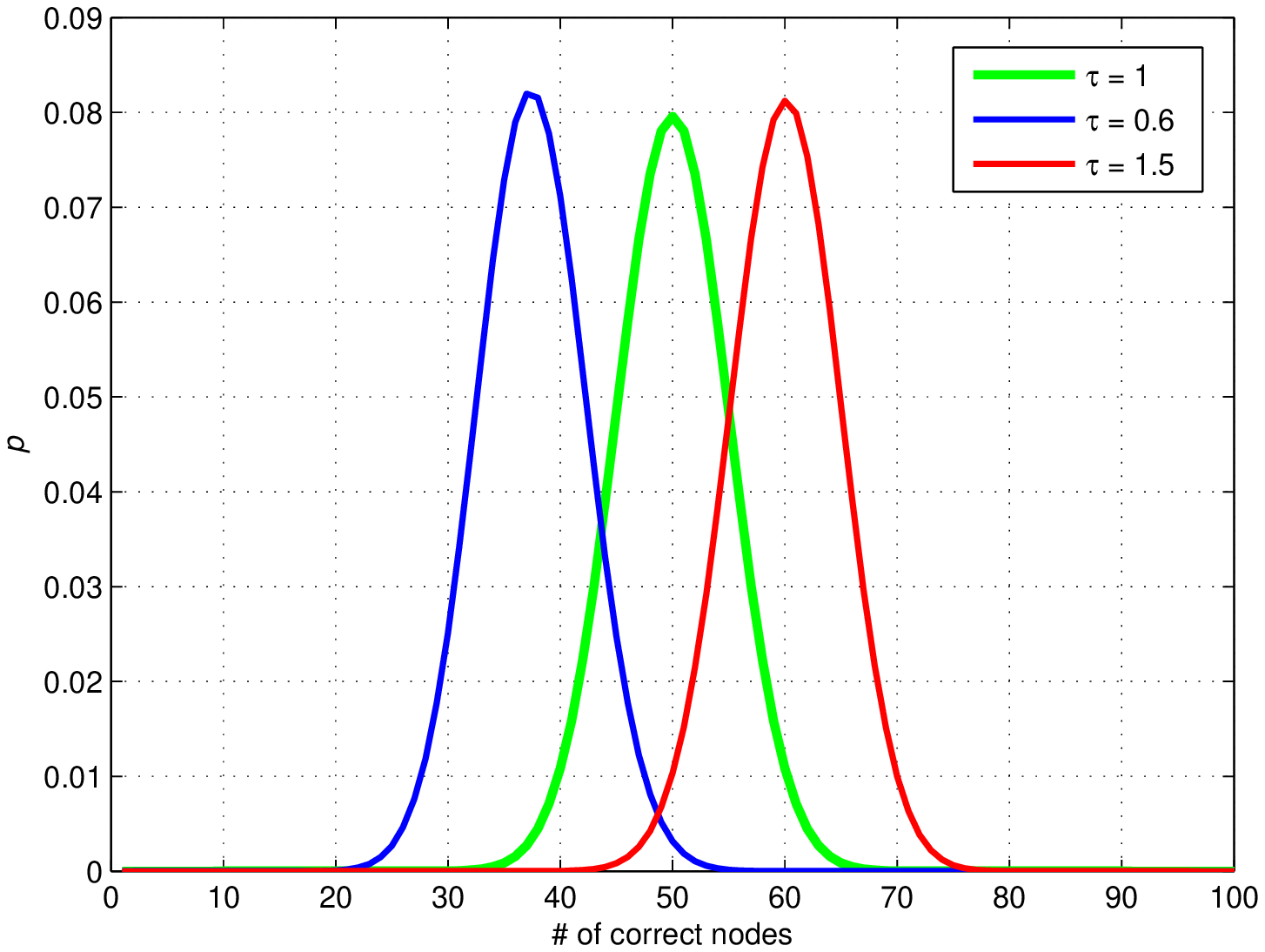}
    \caption{Stationary distribution $\pi$ with $N=100$, $\lambda=1$,
    and $\tau=0.6,1,1.5$. The $y$-axis is the probability density corresponding
    to a certain number of correct nodes. Since only the product
    $\lambda\tau$ matters, we choose the values of $\lambda$ and $\tau$ arbitrarily
    without a dimension.}\label{fig:sd}
   \end{center}
\end{figure}

Therefore, the transition matrix of the subordinated chain
$\{\hat{X}_{n}\}_{n\geq0}$ is as follows:
\begin{displaymath}
\mathbf{P}=\left[
\begin{array}{lllllll}
s_{0} & \nu_{0} &&&&& \\
\mu_{1} & s_{1} & \nu_{1} &&&& \\
& \ddots & \ddots & \ddots && \\
&& \mu_{i} & s_{i} & \nu_{i} & \\
&&& \ddots & \ddots & \ddots  \\
&&&&& \mu_{N} & s_{N} \\
\end{array}
\right]
\end{displaymath}
where $i$ is the number of correct nodes in the system,
$\mu_{i}=\frac{i}{N\tau(\lambda+\tau^{-1})}$ and
$\nu_{i}=\frac{(N-i)\lambda}{N(\lambda+\tau^{-1})}$ represent the
transitions resulting from a compromising and a refreshing,
respectively, and
$s_{i}=\frac{N-i}{N\tau(\lambda+\tau^{-1})}+\frac{i\lambda}{N(\lambda+\tau^{-1})}$
expresses those fruitless operations. One can easily see that this
is a birth-and-death process in continuous time with reflecting
barriers at 0 and $N$ \cite{Bremaud99}. The chain
$\{\hat{X}_{n}\}_{n\geq0}$ is \textit{irreducible} (i.e., every
state is reachable from all other states) and \textit{positive
recurrent} (i.e., the system does not freeze at some states). It has
the following stationary distribution (the detailed computation is
omitted):
\begin{eqnarray}
\pi_{0}&=&\left\{1+\frac{\nu_{0}}{\mu_{1}}+
\frac{\nu_{0}\nu_{1}}{\mu_{1}\mu_{2}}+\cdots+\frac{\nu_{0}\nu_{1}\cdots\nu_{N-1}}{\mu_{1}\mu_{2}\cdots\mu_{N}}\right\}^{-1} \\
\pi_{i}&=&\pi_{0}\frac{\nu_{0}\nu_{1}\cdots\nu_{i-1}}{\mu_{1}\mu_{2}\cdots\mu_{i}}
\end{eqnarray} Note that this is also the stationary distribution of
$\{X(t)\}_{t\geq0}$. It has the following properties:
\begin{itemize}
\item The system can rarely be free either of correct nodes
($X(t)=0$) or of compromised nodes ($X(t)=N$), because both
$\pi_{0}$ and $\pi_{N}$ vanish with increasing $N$.

\item The most likely state (i.e., $\arg\max_{i}\pi_{i}$) lies
between $0$ and $N$; it depends on the magnitude of $\lambda$ and
$\tau^{-1}$. The larger the value of $\lambda\tau$ (the ratio
between the rate of refreshing and that of compromising) is, the
closer is this state to $N$.
\end{itemize} These two properties can be easily observed in Fig.~\ref{fig:sd}.
 It shows that even if the sink is more efficient
than the adversary ($\lambda\tau=1.5$, the red curve), there are
still approximately 40$\%$ compromised nodes.

Now, we evaluate the probability of having at least one correct node
re-encrypting the data on a routing path of length $L$ from a source
to the adversary. Let a random variable $Y$ be the number of correct
nodes re-encrypting the data and hence
\begin{eqnarray}
\textstyle Y=\sum_{m=1}^{M}\Omega_{m}~~~~M\leq L
\end{eqnarray} where $M$ is the random variable representing the
number of nodes that re-encrypt the data and $\{\Omega_{m}\}$ are
i.i.d.  Bernoulli variables indicating the state of each of the $M$
nodes ($\Omega_{m}=1$ if correct and 0 otherwise). We want to
calculate $\mathrm{P}\{Y>0\}=1-\mathrm{P}\{Y=0\}$, the
\textit{success probability} (in the sense that GossiCrypt
successfully provides confidentiality). To this end, we make use of
the generating function $g_{Y}(z)$ of $Y$, because
$\mathrm{P}\{Y=0\}=g_{Y}(0)$ and, by the rule of \textit{random sum
of i.i.d. variables} \cite{Bremaud99},
$g_{Y}(z)=g_{M}(g_{\Omega}(z))$. Therefore,
\begin{eqnarray}
\hspace{-1.5em}\mathrm{P}\{Y=0\}&=&g_{M}(g_{\Omega}(0)) \nonumber \\
&=&\mathrm{E}_{M}[\mathrm{P}\{\Omega_{0}=0\}^{m}] \nonumber \\
&=&\sum_{m=1}^{L}\mathrm{P}\{\Omega_{0}=0\}^{m}\binom{L}{m}q^{m}(1-q)^{L-m}
\end{eqnarray} Given the stationary distribution $\pi$ of
$\{X(t)\}_{t\geq0}$,
\begin{eqnarray}
\mathrm{P}\{\Omega_{0}=0\}&=&\sum_{i=0}^{N}\mathrm{P}\{\Pi, X(t)=i\} \nonumber \\
&=&\sum_{i=0}^{N}\frac{N-i}{N}\pi_{i}
=\frac{N-\mathrm{E_{\pi}}(X)}{N}
\end{eqnarray} where $\Pi$ is the event of picking a node within $N-i$ compromised
ones. We illustrate the success probability $\mathrm{P}\{Y>0\}$
under different values of $L$ and $q$ in Table~\ref{tab:sp},
assuming $N=100$, $\lambda=1$, and $\tau=1.5$. One might think the
case where $\mathrm{P}\{Y>0\}=0.8258$ (for $L=5$ and $q=0.5$) is an
unfavorable bet for the legitimate user (because the adversary is
able to decrypt the data with probability 0.1742); the adversary,
however, gains nothing from this. To understand this point, we refer
again to Fig.~\ref{fig:MobileAdv}. Since what the adversary might
decrypt (with probability 0.1742) is just a snapshot, the
probability of observing the whole data history goes to zero (the
probability of obtaining three snapshots is already very low:
$0.1742^{3}=0.0053$). Note that we take for granted that the events
of decrypting two different snapshots are independent; this is
guaranteed by the coin flipping procedure even if two snapshots are
transmitted through the same routing path.
\begin{table}[htb] \center\small
\begin{tabular}{|l||c|c|c|c|c|}
\hline \vspace{-1ex}~~~$q$ & 0.5 & 0.6 & 0.7 & 0.8 & 0.9  \\
$L$ & & &&&\\ \hline
\hline 5  & 0.8258 & 0.8875 & 0.9303 & 0.9590 & 0.9773 \\
\hline 6  & 0.8772 & 0.9273 & 0.9591 & 0.9783 & 0.9894 \\
\hline 7  & 0.9134 & 0.9531 & 0.9760 & 0.9886 & 0.9950 \\
\hline 8  & 0.9390 & 0.9697 & 0.9859 & 0.9940 & 0.9977 \\
\hline 9  & 0.9570 & 0.9804 & 0.9917 & 0.9968 & 0.9989 \\
\hline 10 & 0.9697 & 0.9873 & 0.9951 & 0.9983 & 0.9995 \\
\hline 11 & 0.9786 & 0.9918 & 0.9871 & 0.9991 & 0.9998 \\
\hline 12 & 0.9849 & 0.9947 & 0.9983 & 0.9995 & 0.9999 \\
\hline
\end{tabular}
\caption{Success probability $\mathrm{P}\{Y>0\}$ under different
values of $L$ (path length) and $q$ (coin flip
probability).}\label{tab:sp}
\end{table}

We analyzed to this point the system state process and the
per-message protection due to GossiCrypt given the path length $L$.
In general, $L$ is a random variable. If we knew its probability
distribution $\mathrm{P}(L)$, the probability of breaking the
confidentiality of a single measurement ($T=t_{0}$) from a given
node ($\Delta=1/N$) would~be
\begin{equation}
\mathcal{F}_{t_{0},\frac{1}{N}}=\mathrm{E}_{L}[1-\mathrm{P}\{Y>0\}]
\end{equation}
What we are interested though, as per our specification, is the
confidentiality with respect to any $\Delta \ge 1/N$, and $T = kt_0$
for integer $k\ge1$. Clearly, it depends on $\mathrm{P}(L)$ that is
a complicated consequence of the relative placement of the sink and
sources, as well as the patterns by which the adversary compromises
nodes and the sink refreshes them. As a result, we proceed without
making an assumption on $P(L)$ and describe the property of
GossiCrypt in an asymptotical sense.

\vspace{1mm} \noindent \textbf{Claim:} \emph{GossiCrypt guarantees
the} $\Delta_{T}$-\textbf{Confidentiality} \emph{property for
$\Delta\geq1/N$ with probability $\mathcal{P}$ (with $N$ being the
system size), and $\mathcal{P}\rightarrow1$ when $T\gg t_{0}$.}

\begin{IEEEproof} As it is at least as hard to breach the
confidentiality of two or more measurements as that of a single one,
it is clear that $\mathcal{F}_{t_{0},\Delta}\leq
\mathcal{F}_{t_{0},\frac{1}{N}}$ for any $\Delta>\frac{1}{N}$. The
strict inequality holds if the events of compromising two or more
measurements are independent. Furthermore, we have that
$\mathcal{F}_{T,\Delta}=(\mathcal{F}_{t_{0},\Delta})^{k}$ for
$T=kt_{0}, k>0$. Therefore,
$\mathcal{P}=1-\mathcal{F}_{T,\Delta}\geq1-(\mathcal{F}_{t_{0},\frac{1}{N}})^{k}\rightarrow1$
if $k\rightarrow\infty$. In other words, as $k$ grows, the
probability of safeguarding the confidentiality of $\Delta$
measurements over a period $T$ goes to one. Literally, if the data
history to be captured is sufficiently long, there is virtually no
opportunity for the adversary to succeed in breaking its
confidentiality.\end{IEEEproof}

As shown in Fig.~\ref{fig:sd}, it is always preferable to have
$\lambda\tau>1$ (although $\lambda\tau<1$ can be compensated by
aggressively setting $q$). This is not hard to achieve because,
whereas the adversary obtains keys via its physical presence, the
key refreshing is performed automatically and remotely. A
conservative way to achieve this is to estimate
$\tau_{\mathrm{min}}$ (the lower bound of $\tau$) and to set
$\lambda>\tau^{-1}_{\mathrm{min}}$. Estimating $\tau$ online can be
preferable. We also note the the convergence of $\mathcal{P}$
persists even if $\lambda\tau<1$ but, of course, with a lower speed.

\subsection{Energy Expenditure} \label{sec:ee}
As we mentioned in Sec.~\ref{sec:in}, applying PKE is an alternative
solution to thwart a parasitic adversary. We will show in this
section that, a sound in theory PKE-based solution is inferior to
GossiCrypt due to the much higher energy expenditure it incurs.

For a quantitative comparison between PKE and GossiCrypt, we make
the following assumptions:
\begin{enumerate}
\item The network size $N<2^{16}$, so node identity $S_i$ needs at most 16
bits. \vspace{-.5ex}
\item Each message has a length of 20 bytes.
\item GossiCrypt makes use of AES-128 encryption.
\item The PKE can either be RSA-1024 or ECC-160.\footnote{Rabin PKE,
in theory, is more efficient than RSA (though the difference can be
as low as one modular multiplication for low RSA exponent
operations) \cite{Menezes97}. However, we are not aware of sensor
network software implementations for Rabin PKE. Moreover, Rabin
appears to be costlier than RSA certain implementations in other
platforms \cite{Crypto++}.}
\item The energy expenditure for transmission is 0.21 $\mu J/\mathrm{bit}$.
\end{enumerate}
The transmission cost refers to MICA2 nodes, and so are the
computation delays for cryptographic operations, and the related
power dissipation, based on available experimental results. Note
that the fourth assumption strongly favors PKE, with its 80-bit
security compared with the AES 128-bit security level. The energy
costs are taken from \cite{PKLife}. Although hardware
implementations could significantly reduce energy consumption for
all primitives \cite{AES004,ECC066,RSA512}, the order of difference
is maintained.

Table~\ref{tab:comp} compares GossiCrypt with two variants of PKE in
terms of computation\footnote{The computational complexity is
measured in different units for symmetric-key and public-key
encryption in \cite{PKLife}. So we need to fix the message size in
order to compare them.} and communication complexity.
\begin{table}[htb] \center\small
\begin{tabular}{|l|c|c|c|}
\hline
 & GossiCrypt & PKE-RSA & PKE-ECC  \\
%\hline Comp. HW & 0.04 $nJ/\mathrm{bit}$ & 0.83 $nJ/\mathrm{bit}$ & 4.12 $\mu J/\mathrm{bit}$  \\
%\hline Comp. & 32.4 $\mu J/\mathrm{msg}$ & 6.60 $mJ/\mathrm{msg}$ & 94.6 $mJ/\mathrm{msg}$  \\
\hline Comp. & 32.4 $\mu J/\mathrm{msg}$ & 14.1 $mJ/\mathrm{msg}$ & 53.4 $mJ/\mathrm{msg}$  \\
\hline       & An increase of $16q$ bits & 1024 bits & $320$ bits\\
       Comm. & per message per hop & per message & per message  \\
\hline
\end{tabular}
\caption{Comparison between GossiCrypt and PKEs.}\label{tab:comp}
\end{table}

We have the following observation on Table~\ref{tab:comp}: First,
the energy expenditure in computation of GossiCrypt at a source node
is 2 to 3 orders of magnitude lower than the those of PKEs. Second,
the energy expenditure in communication of GossiCrypt for each node
en-route remains lower than those of PKEs up to 10$q^{-1}$ (for
PKE-ECC) and 54$q^{-1}$ (for PKE-RSA) hops (note that $q<1$).

It is clear that the communication cost of GossiCrypt is lower than
that of PKE-ECC below 10$q^{-1}$ hops and that of PKE-RSA below
54$q^{-1}$ hops. We assume the scale of the WSN meets these criteria
and we only compare the computation cost below. Note that assuming
20 bytes message actually favors PKE-ECC, whose cost would be
doubled if, for example, the message were one byte longer.

The additional computation cost for GossiCrypt compared with PKE
stems from key refreshing; we denote it as $c_{\mathrm{refresh}}$.
Based on the analysis in Sec.~\ref{sec:sa}, let us assume refresh
rate equal to the adversary compromise rate (i.e., $\lambda\tau =
1$). For $T=k t_0$, let $\tau =T/k$ as per the definition of the
parasitic adversary, or in other words, the adversary compromises
one node per measurement period $t_0$. Then, for a (sub-)network of
$N$ nodes among which the sink picks randomly, each node will be
refreshed on the average once every $N$ measurement periods. The
advantage for GossiCrypt per source node is approximately the ratio
of $\frac{N\times c_{\mathrm{GC}}+c_{\mathrm{refresh}}}{N\times
c_{\mathrm{PKE}}}\approx\frac{N+1}{N}\frac{c_{\mathrm{GC}}}{c_{\mathrm{PKE}}}$
without public-key encryption (as $c_{\mathrm{GC}}\approx
c_{\mathrm{refresh}}$) or $\approx
\frac{1}{N}\frac{c_{\mathrm{refresh}}}{c_{\mathrm{PKE}}}$ with
public-key encryption (as
$\frac{c_{\mathrm{GC}}}{c_{\mathrm{PKE}}}\ll1$), where
$c_{\mathrm{GC}}$ and $c_{\mathrm{PKE}}$ are the computation costs
for GossiCrypt and PKEs, respectively, given in
Table~\ref{tab:comp}.

As the advantage of GossiCrypt over PKEs is \textbf{tremendous}
without public-key encryption, we only consider the key refreshing
with ECC-based public-key encryption. In this case, the cost of
refreshing is dominated by one ECC encryption, thus
$\frac{c_{\mathrm{refresh}}}{c_{\mathrm{PKE}}}\approx1$. Therefore,
the ratio $\frac{1}{N}\frac{c_{\mathrm{refresh}}}{c_{\mathrm{PKE}}}$
decreases as $N$ grows, thus making GossiCrypt increasingly
advantageous. For example, if $N=100$, GossiCrypt can be \textbf{100
times} less costly then PKE-ECC. For PKE-RSA,
$c_{\mathrm{refresh}}\approx3c_{\mathrm{PKE}}$ and GossiCrypt is
still 33 times less costly. However, the very high communication cost
of PKE-RSA is a significant disadvantage that makes PKE-RSA
infeasible.

The comparison above might seem unfair, as one could argue that
using PKE on a per-message basis is not necessary; for example, PKE
could be used only to ``transport'' a symmetric key from each source
sensor node to the sink. Then, such end-to-end symmetric keys could
be the only ones to be used to encrypt once data measurements only
at the source. Clearly, such symmetric keys would be used for
numerous subsequent data messages, followed by a new key transport.
However, as we emphasized in Sec.~\ref{sec:in}, such conventional
key refreshing does not fully thwart the parasitic adversary:
between two refreshing events, the adversary would still be fully
capable of compromising nodes and hence decrypting their data.
Therefore, to reach the security level achieved by GossiCrypt,
conventional key refreshing has to be performed frequently for
almost all nodes. Given our assumption that the adversary
compromises one node per measurement period $t_0$, without
GossiCrypt all $N$ (symmetric) keys would have to be refreshed every
$t_0$. Since $c_{\mathrm{refresh}}\geq c_{\mathrm{PKE}}$ in general,
it would be more efficient to just use PKE on a per-message basis.

\section{Experiment Results}\label{sec:sim}

\begin{figure}[htb]
   \begin{center}
        \includegraphics[width=0.8\columnwidth]{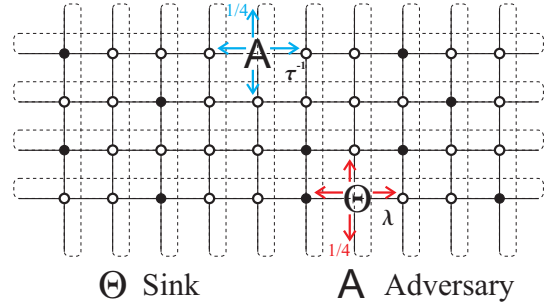}
    \caption{Simulation settings.}\label{fig:sim}
   \end{center}
\end{figure}

We perform simulations in Matlab. We only simulate the operations of
GossiCrypt without taking the MAC/PHY effects into account. We
assume a grid network where nodes appear on a
$\sqrt{N}\times\sqrt{N}$ square lattice. The movements\footnote{We
note that the sink may make a virtual movement by simply changing
the target of the key refreshing protocol, but the adversary has to
always physically move to a node to launch its attack.} of both the
sink and the adversary follow a 2D random walk: they take identical
probability $1/4$ in choosing one direction out of four
possibilities. The intervals between two successive events of moving
follow exponential distributions with mean $\lambda^{-1}$ and $\tau$
for the sink and the adversary, respectively. We assume $N=100$,
$\lambda=1$, and $\tau=1.5$. To remove the boundary effect, we
project the lattice on a torus, i.e., moving out of the one side of
the lattice leads to entering on the opposite side. We illustrate
these settings in Fig.~\ref{fig:sim}.

\begin{figure}[htb]
   \begin{center}
        \includegraphics[width=.85\columnwidth]{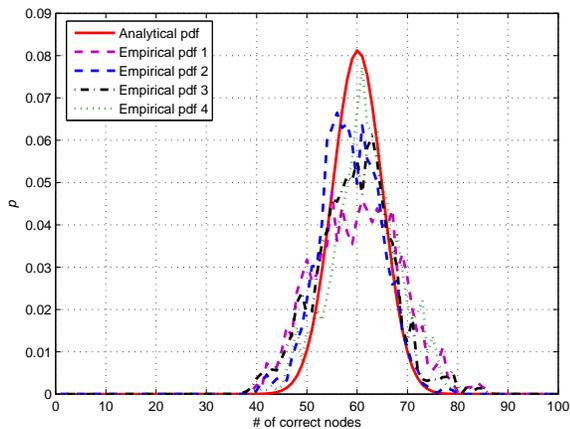}
        \caption{Stationary distributions of the number of correct nodes.}\label{fig:simpdf}
   \end{center}
\end{figure}

Since the stochastic process described above can be proved to be
aperiodic and positive recurrent, all the states are
ergodic~\cite{Bremaud99}. Therefore, we can use statistics over time
to characterize the stationary distribution. We run each simulation
for 11000 transitions and truncate the first 1000 points (which are
in transient phase), such that the results are measured in steady
state. Fig.~\ref{fig:simpdf} shows the comparison between four
empirical stationary distributions resulting from four simulation
runs and the analytical one obtained in Sec.~\ref{sec:sa},
It is clear that the analytical results describe the stationary
regime of the system very well.

\begin{figure}[htb]
   \begin{center}
        \includegraphics[width=.85\columnwidth]{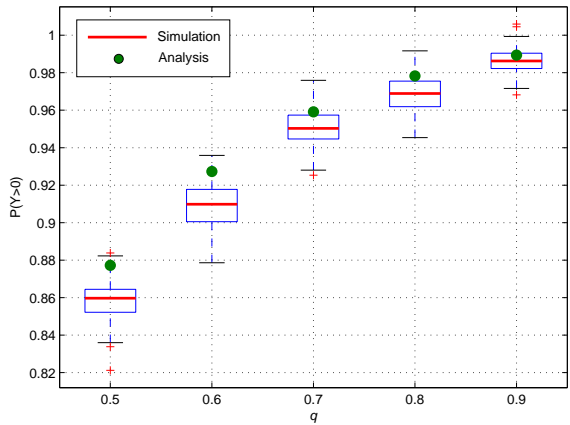}
    \caption{Successful probability $\mathrm{P}\{Y>0\}$ as function of the GossiCrypt parameter $q$.}\label{fig:sprob}
   \end{center}
\end{figure}

Based on these statistics, we can again verify the success
probability $\mathrm{P}\{Y>0\}$ by randomly choosing routing paths
between nodes and the adversary. For brevity, we only illustrate the
case with $L=6$ in Fig.~\ref{fig:sprob} (showing the medians and
$95\%$ quantiles) and compare the results with the analytical ones
shown in Table~\ref{tab:sp}.
The comparison shows that the analytical results are a bit
overoptimistic, but the differences with the experiment results are
negligible.

\begin{figure}[htb]
   \begin{center}
        \includegraphics[width=.85\columnwidth]{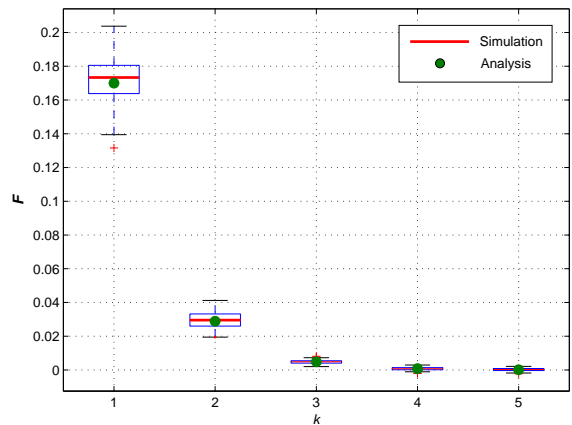}
    \caption{The probability of breaking the confidentiality of $k$
    measurements from a given node $\mathcal{F}_{kt_{0},\frac{1}{N}}$ as function of $k$.}\label{fig:deltaconf}
   \end{center}
\end{figure}

Finally, we verify our claim that GossiCrypt guarantees the
$\Delta_{T}$-\textbf{Confidentiality} property with probability
almost one when $T=kt_{0}$ is sufficiently long. To this end, we
randomly pick two nodes on the grid and consider one as the source
and the other as the data collector. By applying GossiCrypt to the
shortest path between the two nodes, we can evaluate the quantity
$\mathcal{F}_{kt_{0},\frac{1}{N}}$ for different values of $k$. As
shown in Fig.~\ref{fig:deltaconf}, this probability converges very
fast to zero with an increasing $k$, according to both simulation
and analytical results. This corroborates our claim that
$\mathcal{P}=1-\mathcal{F}_{T,\Delta}\rightarrow1$.

To summarize our results in the analysis of Sec.~\ref{sec:sa} and
the experiments of this section: we showed that, for any protocol-
or application-specific objective $\Delta\ge1/N$, the
confidentiality of the sensed data can be safeguarded with
probability almost equal to one. Although this seems to require that
a sufficiently high number of measurements (or equivalently long
period $T$) are of interest, analytic and experimental values show
that even very short sequences (e.g., $T=5t_0$) of measurements
originating from a single source node can be protected with
probability fast approaching one. This is achieved thanks to the
GossiCrypt en-route encryption, resulting in particularly robust
operation even when approximately $40\%$ of the nodes are
compromised by the adversary (as shown by Fig.~\ref{fig:simpdf}).

\section{Discussion}\label{sec:disc}
As described in Sec.~\ref{sec:kr}, the key refreshing protocol does
not provide reliable communication. Hop-by-hop re-transmissions can
remedy transient packet loss, but it may still be possible that a
key refreshing message sent from a node $S_i$ to the sink $\Theta$
is lost. In that case, $\Theta$ would be unable to decrypt messages
$S_i$ encrypts with the new (``refreshed'') symmetric key. A
multi-round sensor node-sink communication protocol, to confirm at
both ends the key refreshing was successful, would not be an option.
This is so as the sink response could single-handedly divulge the
node that performed the refreshing, and thus enable to adversary to
target the node and re-compromise it to obtain the new key. As a
result, we propose here a straightforward solution: to add limited
redundancy only for the infrequent key refreshing messages. One
option is to let $S_i$ repeat the same message a few times; in the
presence of benign faults the probability of successful reception
will be practically one with a few repetitions. Depending on the
underlying networking protocol, if, for example, nodes form a
directed acyclic graph rather than a tree, each node could transmit
key refreshing message replicas to different neighbors and thus
across different paths. Of course, adding redundancy leads to higher
overhead; for example, instead of transmitting one key refreshing
message over $N$ nodes per
$t_0$ seconds, $r$ would be transmitted, but GossiCrypt is still advantageous as $r\ll N$. %% Shall we make a statement here?
%% (1) perhaps, we can make a positive statement here in the place of your phrase, even if sth is such and such,
%%     for some r=2,3,4 the improvement over traditional PKE... is r/N * (c_.../c_...)
% -- "the tension between reliability and overhead is irreconcilable in distributed
% -- settings."
%% (1.b) There is this paper by Heidelman that looks at e-t-e vs. hop-by-hop communication...
%% (2) Shall we put here a comment that "we do not consider anonymous communication"?
%% (3) Note: I thought over the coding stuff... I am not sure what is the impact of packets that
%% are distinguishable due to variable sizes... perhaps a topic for later work, unless you think it is worth mentioning

The impact of active adversaries is discussed next. After
compromising a key, they can impersonate $S_i$, and invoke a fake key
refreshing.\footnote{Public key cryptography (e.g., digital
signatures generated by a source node $S_i$) would not be
advantageous: the private key of $S_i$ can be obtained by an
adversary that physically compromises $S_i$.} The adversary could
then establish a new shared key with the sink. At first, the
impersonating adversary would be constrained in terms of where to
invoke the fake refreshing from, as the $S_i$-to-$\Theta$ path is
essentially accumulated in the key refreshing message. Independently
of that, however, once the actual key refreshing occurs, $S_i$ will
operate with a different key from its impostor. The unobtrusive
adversary cannot prevent $S_i$ from launching a key refreshing
protocol, and it cannot upload its own ``new'' key to $S_i$. As a
result, even if the adversary controls the $S_i$-to-$\Theta$
communication, it can at most deny data collection from $S_i$. But
the active adversary would fail to obtain the data $S_i$ reports
encrypted with the actual new key, unless it re-compromises
physically $S_i$.

%Consequently, $\Theta$ will detect the inconsistency at a later
%interaction with $S_i$,
%% This to avoid the critisism of a new dos or nuisance attack, that
%% sends the administrators out in the field...

%% ", while preventing the reception of messages from
%% the actual $S_i$." -- I took this out because you talk about a later

As a follow-up work, we intend to consider specific instantiations
of WSNs, e.g., network sizes and topologies, data extraction and key
refreshing methods, and value ranges for other system
characteristics such as $\delta$, $T$, and $\Delta$, and $\tau$ and
$\lambda$. Extending our work in this way, through analytical and
experimental means, would allow us to investigate a number of
interesting questions. For example, postulate fine-grained claims
conditional on specific networks, revealing design trade-offs due to
the relative roles of $\Delta$ and $T$. Or, identify the right
``mix'' of symmetric- and public-key based key refreshing
techniques, as a function of the adversary presence, to evaluate the
trade-off of effectiveness for cost.

\section{Conclusion} \label{sec:con}

As security becomes an important requirement for WSNs, the salient
characteristics of WSNs clue the more relevant threats and types of
exploit to thwart with practical defense mechanisms. With this
consideration in mind, we identify here a novel threat, a parasitic
adversary, targeting exactly the most valuable asset of a WSN, its
measurements. The parasitic adversary is a practical and realistic
threat because of (i) its well-aimed exploit, unauthorized access to
WSN data, (ii) its well-chosen methods, targeting at the weakest
system point, the low physical sensor node protection, and (iii) its
resource constraints and ``low-profile" operation.

The second and main contribution of this paper is GossiCrypt, a
scheme to ensure WSN data confidentiality. GossiCrypt's two building
blocks are a probabilistic en route encryption of the data towards
the sink and a key refreshing mechanism, both leveraging on the scale
of WSNs. The former relies on very simple key management assumptions,
it is simple in operation. The latter reverses the impact of the
physical compromise of sensor nodes.

%The resilience of our scheme is enhanced by, even though it does not
%depend upon, the use of mobile sinks for data collection.

Our evaluation shows that GossiCrypt can prevent the breach of WSN
confidentiality in a wide range of settings. Even though the
adversary could obtain solitary or sparse measurements, our analysis
and simulations show that GossiCrypt prevents the compromise of a
meaningful set of measurements over a period of time with
probability going to one. The most intriguing feature of GossiCrypt
lies in its ability of defending the WSN data confidentiality with
simple and low-cost mechanisms. We believe that such approaches that
leverage on the WSN characteristics, rather than imitating iron-clad
approaches from other distributed computing paradigms, can be
effective in addressing security challenges for wireless sensor
networks.

\bibliographystyle{plain}
\bibliography{GossiCrypt}

\end{document}